\documentclass[12pt]{article}
\usepackage{graphicx}

\title{Effect of optical purity on phase sequence in antiferroelectric liquid
crystals}
\author{Nata\v{s}a Vaupoti\v{c}$^{1,2}$ and Mojca \v{C}epi\v{c}$^{1,3}$ \\
$^{1}${\small Institut Jozef Stefan, Ljubljana, Slovenia}\\
$^{2}${\small Department of Physics, Faculty of Education,
University of
Maribor, Maribor, Slovenia}\\
$^{3}${\small Department of Physics, Faculty of Education,
University of Ljubljana, Maribor, Slovenia}}
\input{tcilatex}
\begin{document}
\maketitle

\begin{abstract}
We use the discrete phenomenological model to study theoretically the phase
diagrams in antiferroelectric liquid crystals (AFLCs) as a function of
optical purity and temperature. Recent experiments have shown that in some
systems the number of phases is reduced if the optical purity is extremely
high. In some materials the SmC$_{A}^{\star }$ phase is the only stable
tilted smectic phase in the pure sample. In the scope of the presented model
this high sensitivity of the phase sequence in the AFLCs to optical purity
is attributed to the piezoelectric coupling which is reduced if optical
purity is reduced. We limit our study to three topologically equal phases -
SmC$^{*}$, SmC$_{\alpha }^{*}$ and SmC$_{A}^{*}$ and show that the reduction
of optical purity forces the system from the antiferroelectric to the
ferroelectric phase with a possible SmC$_{\alpha }^{\star }$ between them.
The effect of the flexoelectric and quadrupolar coupling is considered as
well. If the phase diagram includes only two phases, SmC$^{\star }$ and SmC$%
_{A}^{\star }$, the flexoelectric coupling is very small. The materials
which exhibit the SmC$_{\alpha }^{\star }$ in a certain range of optical
purity and temperature, can be expected to have a significant flexoelectric
coupling that is comparable with the piezoelectric coupling. And finally,
when temperature is lowered the phase sequence SmA $\rightarrow $ SmC$%
_{\alpha }^{\star }$ $\rightarrow $ SmC$^{\star }$ $\rightarrow $ SmC$%
_{A}^{\star }$ is possible only in materials in which quadrupolar coupling
is very strong.
\end{abstract}

\maketitle

\section{Introduction}

Antiferroelectric liquid crystals (AFLCs) were discovered more than 15 years
ago when macroscopic properties of mixtures of left- and right-handed
MHPOBC, nowadays known as a prototype antiferroelectric liquid crystal, were
studied \cite{chandani}. Scientists were surprised at the rich variety of
phases found in these systems and at the influence that the optical purity
has on the phase sequence. There is a number of phenomena found in these
complex systems \cite{exp1}-\cite{Mach} and the macroscopic properties of
some of the phases were explained only recently by the theoretical modelling
of the microscopic structure \cite{mojcajcp}.

In chiral antiferroelectric liquid crystals there always exists at least one
phase with antiferroelectric properties, the SmC$_{A}^{*}$ phase \cite
{fukudaover}. Due to the oppositely (anticlinically) tilted elongated
molecules in the neighboring layers, the piezoelectrically induced
polarization is cancelled out and the system behaves antiferroelectrically
in the external electric field. The structure is additionally helicoidally
modulated by a double helix formed from anticlinically tilted layers. The
length of the pitch depends on optical purity and it is infinite in racemic
mixtures. The synclinically tilted ferroelectric SmC$^{*}$ phase is also
often found in AFLCs. It is stable at higher temperatures than the SmC$%
_{A}^{*}$ phase. In the temperature range between the SmC$_{A}^{*}$ and the
SmC$^{*}$ phase two other phases with a short modulation over 3 and 4 layers
and long helicoidal modulation over at least few hundred of layers can exist
\cite{Mach}. They are called SmC$_{FI,I}^{*}$ and SmC$_{FI,II}^{*}$,
respectively. In optically very pure samples the 4-layer phase is sometimes
stable within the same temperature range as the SmC$^{*}$ phase \cite
{ewa-mhpobc}. Directly below the transition to the tilted phase, the phase
called SmC$_{\alpha }^{*}$ can be stable. This phase is topologically equal
to the SmC$^{*}$ phase but the period of the helicoidal modulation extends
over a few layers only. The phase transition between the SmC$_{\alpha }^{*}$%
\ and SmC$^{*}$ is of an isostructural type and can be recognized by
differential calorimetry measurements only if the changes of the properties
like the length of the modulation period or the tilt magnitude is abrupt
\cite{Garland}. More exact measurements like simultaneous measurements of
optical rotatory power and the tilt \cite{skarabot} have shown that the
phase exist also at lower optical purities but transforms continuously to
the structure of the SmC$^{*}$ phase with a very rapid changes of modulation
period within an extremely narrow temperature range (a few mK) \cite
{skarabot1}.

Optical purity has a strong influence on the phase sequence. It was
generally believed that the number of phases increases with increasing the
optical purity. However, recent experiments \cite{takezoe,lagerwalls} has
shown that in some systems the opposite is true, i.e. the number of phases
is reduced if the optical purity is extremely high. In some materials the SmC%
$_{A}^{\star }$ phase is the only stable tilted smectic phase in the pure
sample.

The motivation of this work was to account for the described properties of
the phase diagram of antiferroelectric liquid crystals within the framework
of the discrete phenomenological model. We limit our consideration to the
phase diagram of three topologically equal phases, also called the clock
phases - SmC$^{*}$, SmC$_{\alpha }^{*}$ and SmC$_{A}^{*}$, that all have
helicoidally modulated structures with pitches of a few hundred layers, of a
few layers, and of approximately two layers, respectively. All three phases
may appear directly below the non-tilted SmA phase. Experimentally three
different phase diagrams have been observed \cite{takezoe}, which are
presented schematically in Fig. \ref{shema}. This high sensitivity of AFLCs
to optical purity has so far been attributed to changes in smectic order
\cite{lagerwalls}. The aim of our study is to show that chiral microscopic
interactions might also be responsible for the observed macroscopic
behavior. The plan of the paper is the following. First we present the
discrete phenomenological model and discuss the theoretical predictions on
how optical purity affects the phase sequences. Then we present
theoretically obtained phase diagrams in which stability regions of phases
are shown as a function of temperature and optical purity. Finally we
discuss the results and draw the conclusions.

\section{Model}

\begin{figure}
\includegraphics[width=0.8\textwidth]{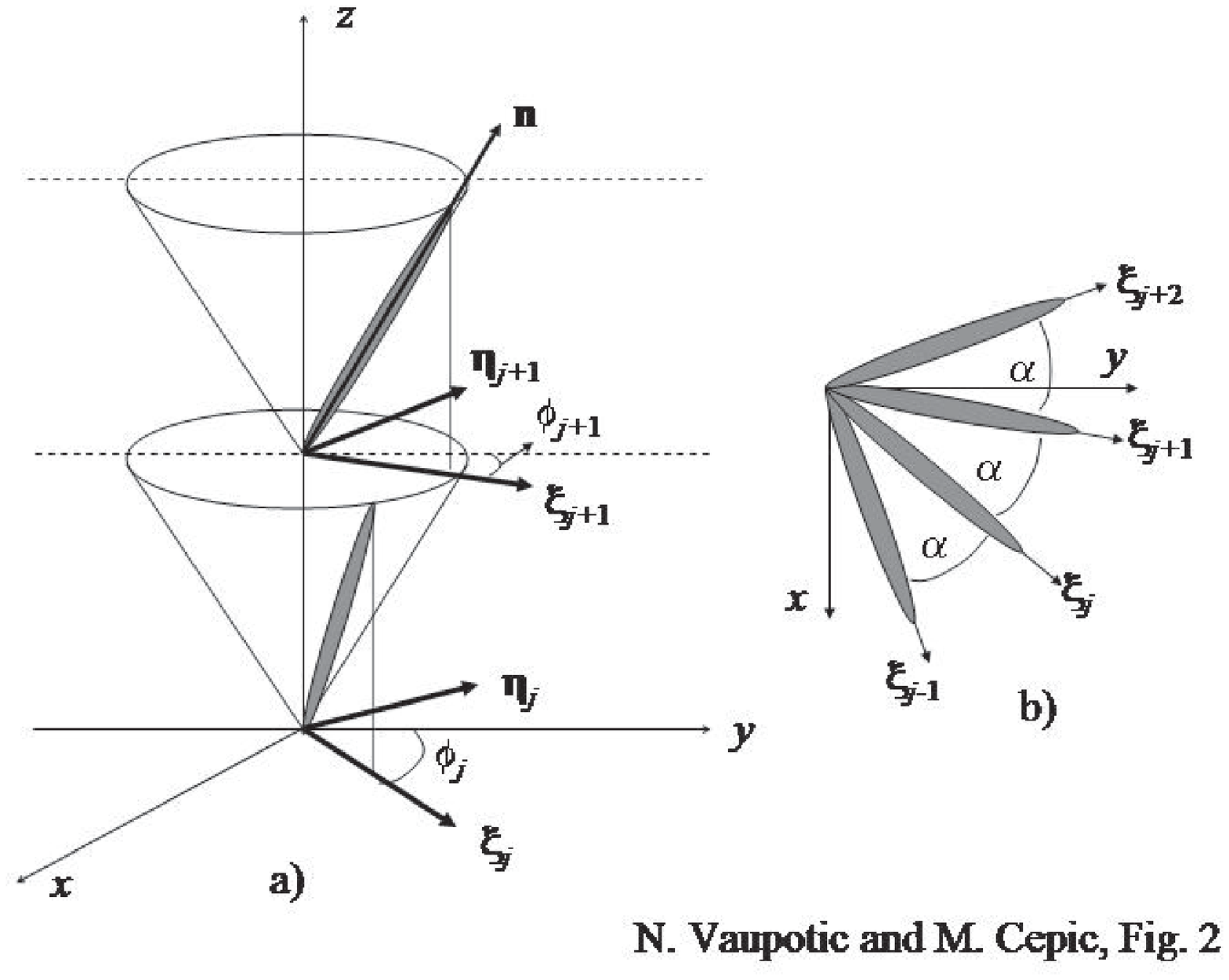}
\caption{The geometry of the problem and the definition of the
order parameters. a) Smectic layers run along the $z$-direction.
Molecules lie on the tilt cone, the magnitude
 of the tilt is $\vartheta $. The system is described by two order parameters:
$\QTR{bf}{\xi }_{j}$ and $\QTR{bf}{\eta }_{j}$. The tilt vector $\QTR{bf}{\xi }_{j}$
points in the direction of the projection of the director $\QTR{bf}{n}$ to the smectic plane
(the $xy$-plane). Its magnitude is $\vartheta $. The phase angle between the tilt vector
 in the $j$-th and the $(j+1)$-th layer is $\phi _{j+1}-\phi _{j}=\alpha $ and it is constant.
In the SmC$^{\star }$ phase $\alpha \approx 0$ and in the SmC$_{A}^{\star }$ phase
 $\alpha \approx \pi $. The polar order parameter $\QTR{bf}{\eta }_{j}$ $\,$lies in the
 smectic plane and it is perpendicular to the tilt vector. b) Top view on the layers.
The elipse now presents the projection of the molecule on the smectic plane.}
\label{orderpar}
\end{figure}

We use the discrete phenomenological model \cite{mojcaprl} and write the
free energy of the system in terms of the tilt vector $\mathbf{\xi }_{j}$
and the polar order parameter $\mathbf{\eta }_{j}$ (see Fig. \ref{orderpar})
in the $j$-th smectic layer. The tilt vector is the projection of the
director $\mathbf{n}$ to the smectic plane and its magnitude is equal to the
tilt $\vartheta $. The spontaneous polarization in the $j$-th layer is
proportional to the polar order parameter: $\mathbf{P}_{j}=P_{0}\mathbf{\eta
}_{j}$, where $P_{0}$ is the polarization of the completely polarly ordered
layer \cite{Photinos}. Because of that the polar order parameter will be
referred to as \textit{polarization}.

The free energy of the smectic system with $N$ layers is:
\begin{eqnarray}
G &=&a\sum_{j=1}^{N}\frac{1}{2}(T-T_{0})\mathbf{\xi }_{j}^{2}+\frac{1}{4}b%
\mathbf{\xi }_{j}^{4}+\frac{1}{2}a_{10}\left( \mathbf{\xi }_{j}\cdot \mathbf{%
\xi }_{j+1}\right) +\frac{1}{2}a_{11}\mathbf{\xi }_{j}^{2}\left( \mathbf{\xi
}_{j}\cdot \mathbf{\xi }_{j+1}\right)   \nonumber \\
&&+\frac{1}{2}f_{1}\left( \mathbf{\xi }_{j}\times \mathbf{\xi }_{j+1}\right)
_{z}+\frac{1}{2}b_{0}\mathbf{\eta }_{j}^{2}+\frac{1}{2}b_{1}\left( \mathbf{%
\eta }_{j}\cdot \mathbf{\eta }_{j+1}\right) +\frac{1}{8}b_{2}\left( \mathbf{%
\eta }_{j}\cdot \mathbf{\eta }_{j+2}\right)   \nonumber \\
&&+c_{p}\left( \mathbf{\eta }_{j}\times \mathbf{\xi }_{j}\right) _{z}+\frac{1%
}{2}\mu \left( \mathbf{\xi }_{j-1}-\mathbf{\xi }_{j+1}\right) \cdot \mathbf{%
\eta }_{j}+\frac{1}{2}b_{Q}\left( \mathbf{\xi }_{j}\cdot \mathbf{\xi }%
_{j+1}\right) ^{2}\ .  \label{enG}
\end{eqnarray}
The significance of the parameters entering the model is discussed in detail
elsewhere \cite{mojcajcp,mojcaprl}, so we make only a short review. The
first two terms describe the intralayer steric and the attractive van der
Waals interactions. The phase transition to the tilted phase in an isolated
layer occurs at temperature $T_{0}$. The parameters $a_{10}$ and $f_{1}$
give the achiral and the chiral part of the van der Waals interaction
between the nearest layers. The fourth term in Eq.(\ref{enG}) describes the
variation of the achiral van der Waals interaction with temperature. The
parameter $a_{11}$ is negative to account for the fact that with reduced
temperature (and thus increased tilt) the antiferroelectric liquid crystal
is always driven to the SmC$_{A}^{*}$ phase. The parameters $b_{0}$, $b_{1}$
and $b_{2}$ give the electrostatic interaction inside the layer, between the
nearest layers and the next nearest layers, respectively. It was shown \cite
{Osipov} that the electrostatic interaction between the next nearest layers
does not significantly affect the structure, therefore in the rest of the
paper we set $b_{2}=0$. The magnitude of the piezoelectric coupling between
the tilt and the polarization is given by $c_{p}$ and the flexoelectric
coupling, i. e. the coupling between the variation of the tilt and the
polarization, is given by the parameter $\mu $. The quadrupolar coupling ($%
b_{Q}$) favors both synclinic and anticlinic ordering in the neighboring
layers \cite{Neal,Ewaprl}, so $b_{Q}$ is negative. All the parameters are
given in the units of K, except the parameter $a$, which has the unit Jm$%
^{-3}$K$^{-1}$.

In this paper we focus only on the clock structures where the polarization ($%
\mathbf{\eta }_{j}$) is perpendicular to the tilt vector and the angle
between the tilt vectors in the neighboring layers is constant and equal to $%
\alpha $. When the free energy (Eq. (\ref{enG})) is minimized with respect
to the magnitude of the polar order parameter $\eta $, we find:
\begin{equation}
\eta =\frac{\left( c_{p}-\mu \sin \alpha \right) \vartheta }{b_{0}+b_{1}\cos
\alpha }\quad .  \label{player}
\end{equation}
The polarization inside one layer thus increases with increasing tilt and as
a result the importance\textbf{\ }of electrostatic interaction increases
when temperature is reduced. The polarization, however, increases also if
the piezoelectric coupling ($c_{p}$) between the neighboring layers
increases. The amount of piezoelectric coupling strongly depends on the
optical purity of the sample. The parameter $x$ is chosen as a measure for
optical purity. In pure samples $x=1$ and in a racemized sample $x=0$. The
piezoelectric parameter can then be written as $c_{p}=xc_{p0}$. The inlayer
polarization (Eq. (\ref{player})) thus decreases with decreasing optical
purity which suggests that materials which exhibit antiferroelectric
structure at high optical purity might be in the ferroelectric state at low
optical purity. Changes in optical purity affect also the chiral part of the
van der Waals interaction ($f_{1}$) which can be rewritten as $f_{1}=xf_{10}$%
.

Elimination of the polar order parameter from Eq. (\ref{enG}) leads to the
free energy expressed only in tilt vectors:
\begin{eqnarray*}
G/a &=&\sum_{j=1}^{N}\frac{1}{2}(T-T_{0})\mathbf{\xi }_{j}^{2}+\frac{1}{4}b%
\mathbf{\xi }_{j}^{4}+\frac{1}{2}a_{11}\mathbf{\xi }_{j}^{2}\left( \mathbf{%
\xi }_{j}\cdot \mathbf{\xi }_{j+1}\right) +\frac{1}{2}b_{Q}\left( \mathbf{%
\xi }_{j}\cdot \mathbf{\xi }_{j+1}\right) ^{2} \\
&&+\frac{1}{2}\sum_{i=1}^{4}\widetilde{a}_{i}\left( \mathbf{\xi }_{j}\cdot
\mathbf{\xi }_{j+i}\right) +\frac{1}{2}\sum_{i=1}^{3}\widetilde{f}_{i}\left(
\mathbf{\xi }_{j}\times \mathbf{\xi }_{j+i}\right) _{z}\ .
\end{eqnarray*}
The electrostatic interaction mediates the indirect interaction among the
layers which extends up to the fourth-nearest layer. The whole set of
effective achiral ($\widetilde{a}_{i}$) and chiral ($\widetilde{f}_{i}$)
parameters is given in \cite{mojcaprl}. The effective parameters between the
nearest and the next nearest layers determine the type of the clock phase
(SmC$_{\alpha }^{\star }$, SmC$_{A}^{\star }$ or SmC$^{\star }$). The
interactions up to the third and the fourth nearest layers are important
only when phases with a short modulation over three and four layers need to
be considered. Although in the numerical calculations we have included the
whole set of the effective parameters, we have checked that $\widetilde{a}%
_{3}$, $\widetilde{a}_{4}$ and $\widetilde{f}_{3}$ actually have no
influence on the phase diagrams. Because of that we discuss in more detail
only the effective parameters between the nearest and the next-nearest
neighboring layers:

\begin{eqnarray}
\widetilde{a}_{1} &=&a_{10}+\left( \frac{c_{p}^{2}}{b_{0}}+\frac{\mu ^{2}}{%
4b_{0}}\right) \frac{b_{1}}{b_{0}}\ ,  \label{a1v} \\
\widetilde{a}_{2} &=&-\frac{c_{p}^{2}}{2b_{0}}\frac{b_{1}^{2}}{b_{0}^{2}}+%
\frac{\mu ^{2}}{2b_{0}}\ ,  \label{a2v} \\
\widetilde{f}_{1} &=&f_{1}+\frac{2c_{p}\mu }{b_{0}}\left( 1+\frac{b_{1}^{2}}{%
4b_{0}^{2}}\right) \ ,  \label{f1v} \\
\widetilde{f}_{2} &=&-\frac{c_{p}\mu }{b_{0}}\frac{b_{1}}{b_{0}}\quad .
\label{f2v}
\end{eqnarray}
The type of the structure (SmC$^{\star }$ or SmC$_{A}^{\star }$) depends
mainly on the sign and the magnitude of the parameter $\widetilde{a}_{1}$.
If it is negative, synclinic ordering between the neighboring layers is
favored, and anticlinic ordering is favored if it is positive. The second
term in $\widetilde{a}_{1}$ (see Eq. (\ref{a1v})) is always positive and its
magnitude increases if $c_{p}$ increases. So, if the direct achiral part of
the van der Waals interaction between the nearest layers is such that it
favors synclinic ordering in the neighboring layers ($a_{10}<0$) the net
effect of all interactions (described by $\widetilde{a}_{1}$) might be such
that the anticlinic ordering between the nearest layers is preferred,
providing that the effect of piezo- and flexoelectric coupling (the second
term in $\widetilde{a}_{1}$ (see Eq. (\ref{a1v}))) is large enough to
overrun the direct van der Waals interaction. This means, that in materials
in which $c_{p0}$ is large compared to $a_{10}$ one can expect that the
reduction of optical purity will drive the material from the
antiferroelectric to the ferroelectric state.

In materials where the effective achiral interaction between the
next-nearest layers ($\widetilde{a}_{2}$) is positive, this interaction
favors anticlinic ordering between the next nearest layers. This ordering is
in contradiction with both the synclinic and the anticlinic ordering in the
nearest layers and it enforces the stability of the SmC$_{\alpha }^{\star }$
phase if the condition $4|\widetilde{a}_{1}|<\widetilde{a}_{2}$ is satisfied
\cite{mojcaprl}. If upon reduction of the optical purity the effective
achiral interaction between the nearest layers changes sign, the stability
condition for the SmC$_{\alpha }^{\star }$ phase is bound to be satisfied in
a certain range of optical purity and the following phase sequence can be
observed when optical purity is reduced: SmC$_{A}^{\star }\rightarrow $ SmC$%
_{\alpha }^{\star }$ $\rightarrow $ SmC$^{\star }$.

\section{Theoretical Phase Diagrams}

\begin{figure}
\includegraphics[width=0.8\textwidth]{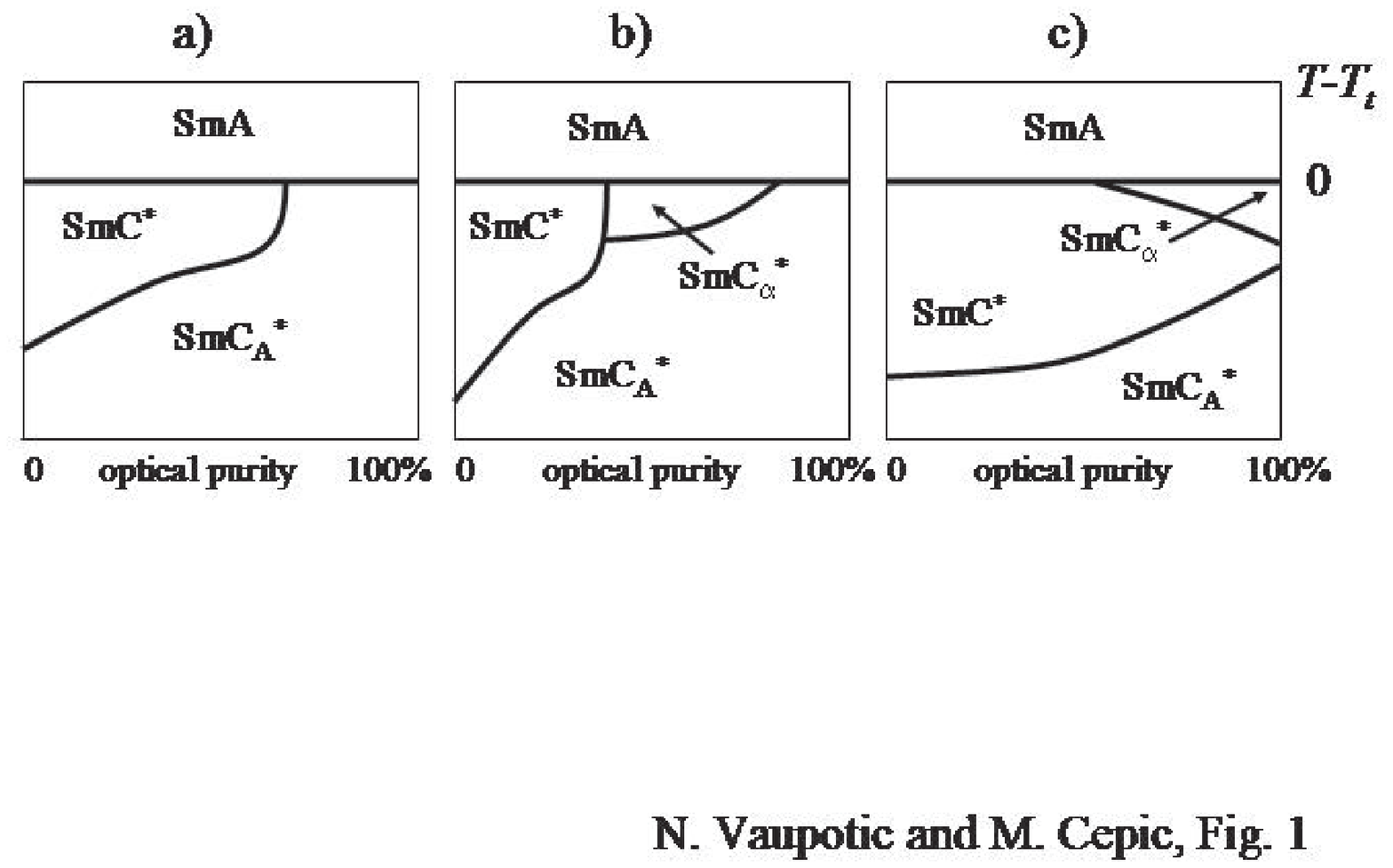}
\caption{Schematic drawings of the three typical phase diagrams observed in AFLCs.
Only the clock structures are shown, the three and four
layer phases are omitted. a) TFMHPOBC, b) TFMHPOCBC and c) MHPOBC.}
\label{shema}
\end{figure}

The aim of this work was to obtain theoretically the phase diagrams that
qualitatively agree with the experimental phase diagrams, shown in Fig. \ref
{shema}. Following that goal we use the model presented above and discuss
the effect that the change in chiral microscopic interactions has on the
phase diagram. In addition we also discuss the effect of the quadrupolar and
the flexoelectric coupling. Finally we show that by looking at the certain
characteristics of the experimental phase diagram one can predict which
macroscopic interactions are most important in a given material.

\begin{figure}
\includegraphics[width=0.8\textwidth]{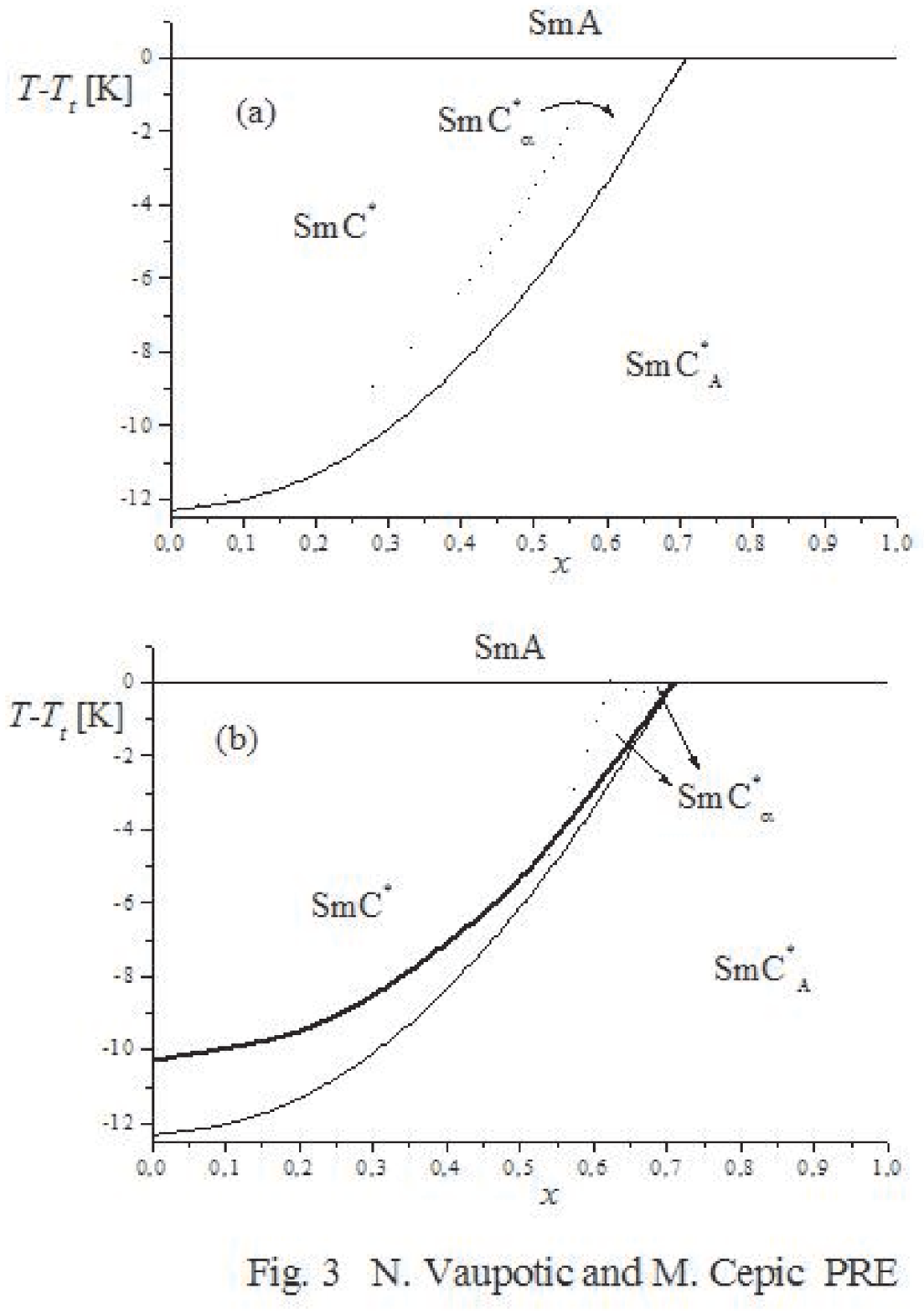}
\caption{Phase diagrams if no flexoelectric coupling is present.
a) $b_Q =0$; without the dotted line: $f_1 =-0.01x$, with the
dotted line: $f_1 = - 0.05x$. b) $f_1 =-0.05x$; thin dotted and
solid line: $b_Q =-1$; thick dotted and solid line: $b_Q =-10$.
Solid and dotted lines denote the first order and the second order
transitions, respectively. The SmC$_{\alpha }^{\star }$ phase is
designated to the region in which $\alpha >0.2$.
  The common set of parameters: $\mu =0$, $c_{p}=4x$,
$b=120$, $a_{10}=-0.4$, $a_{11}=4$, $b_{0}=2$ and $b_1 =b_0 /10$. All the
parameters are given in the units of Kelvin. $T_{t}$ is the transition temperature to the
tilted phase.}
\label{mu0}
\end{figure}

First we focus on the effect of the piezoelectric coupling ($c_{p}$) and set
the flexoelectric ($\mu $) and the quadrupolar coupling ($b_{Q}$) to zero.
In this case only three effective interlayer parameters ($\widetilde{a}%
_{1}=a_{10}+c_{p}^{2}b_{1}/b_{0}^{2}$, $\widetilde{a}%
_{2}=-c_{p}^{2}b_{1}^{2}/(2b_{0}^{3})$ and $\widetilde{f}_{1}=f_{10}x$) are
different from zero. There is no frustration between the nearest and the
next nearest layer interactions ($\widetilde{a}_{2}$ is negative) so a very
simple phase diagram, containing only the SmC$_{A}^{\star }$ phase and the
SmC$^{\star }$ phase, is expected. When the chiral part ($f_{10}$) of the
van der Waals interaction is very small the phase diagram is indeed simple
(see Fig. \ref{mu0}a). This diagram qualitatively agrees with the phase
diagram in Fig. \ref{shema}a. However, if the chiral part of the van der
Waals interaction ($f_{10}$) is increased, the helicoidal modulation can
become so large, that the structure might be recognized as the SmC$_{\alpha
}^{*}$ phase. The phase angle increases continuously when temperature is
lowered. In Fig. \ref{mu0}a we have shown the region of optical purity and
temperature in which the phase angle $\alpha $ is greater than 0.2. This
modulation is already so large that the phase might be recognized as the SmC$%
_{\alpha }^{*}$ phase.

Next we include the quadrupolar coupling ($b_{Q}$), which tends to stabilize
synclinic or anticlinic phases. The presence of small quadrupolar coupling
reduces the region of stability of the SmC$_{\alpha }^{\star }$ phase
significantly as seen in Fig. \ref{mu0}b. This phase diagram qualitatively
agrees with the phase diagram shown in Fig. \ref{shema}b. At even larger
quadrupolar coupling the SmC$^{\star }$ phase 'grows' into the region of the
SmC$_{\alpha }^{\star }$ phase and the SmC$_{\alpha }^{\star }$ phase is
almost expelled from the phase diagram as seen in Fig. \ref{mu0}b.

The diagrams shown in Fig. \ref{mu0} can be used to predict how a phase
diagram for the material which has larger, or lower piezoelectric coupling
looks like. If, for example, the piezoelectric coupling $c_{p0}$ is reduced
by a factor of 1.5, the new phase diagram would be obtained by a simple
rescaling of the optical purity $x$, which should be multiplied by the same
factor, i.e. by 1.5. So, in this case the region of stability of the SmC$%
^{\star }$ phase would extend to higher optical purity, in the chosen
example to $x>1$. In the case of the very strong quadrupolar coupling the
rescaled phase diagram of Fig. \ref{mu0}b qualitatively agrees with the
phase diagram in Fig. \ref{shema}c.

Let us emphasize that when the flexoelectric coupling is negligibly small
the SmC$_{\alpha }^{*}$ phase is obtained solely due to the strong chiral
twist which results from the chiral part of the van der Waals interaction.
We believe, however, that the true origin of the SmC$_{\alpha }^{\star }$
phase is in the frustration between the nearest and the next nearest layer
achiral interactions ($\widetilde{a}_{2}>0$). This frustration is possible
only in materials in which flexoelectric coupling ($\mu $) is important. So
in the rest of the Section we discuss the effect of the flexoelectric
coupling in more detail.

\begin{figure}
\includegraphics[width=0.8\textwidth]{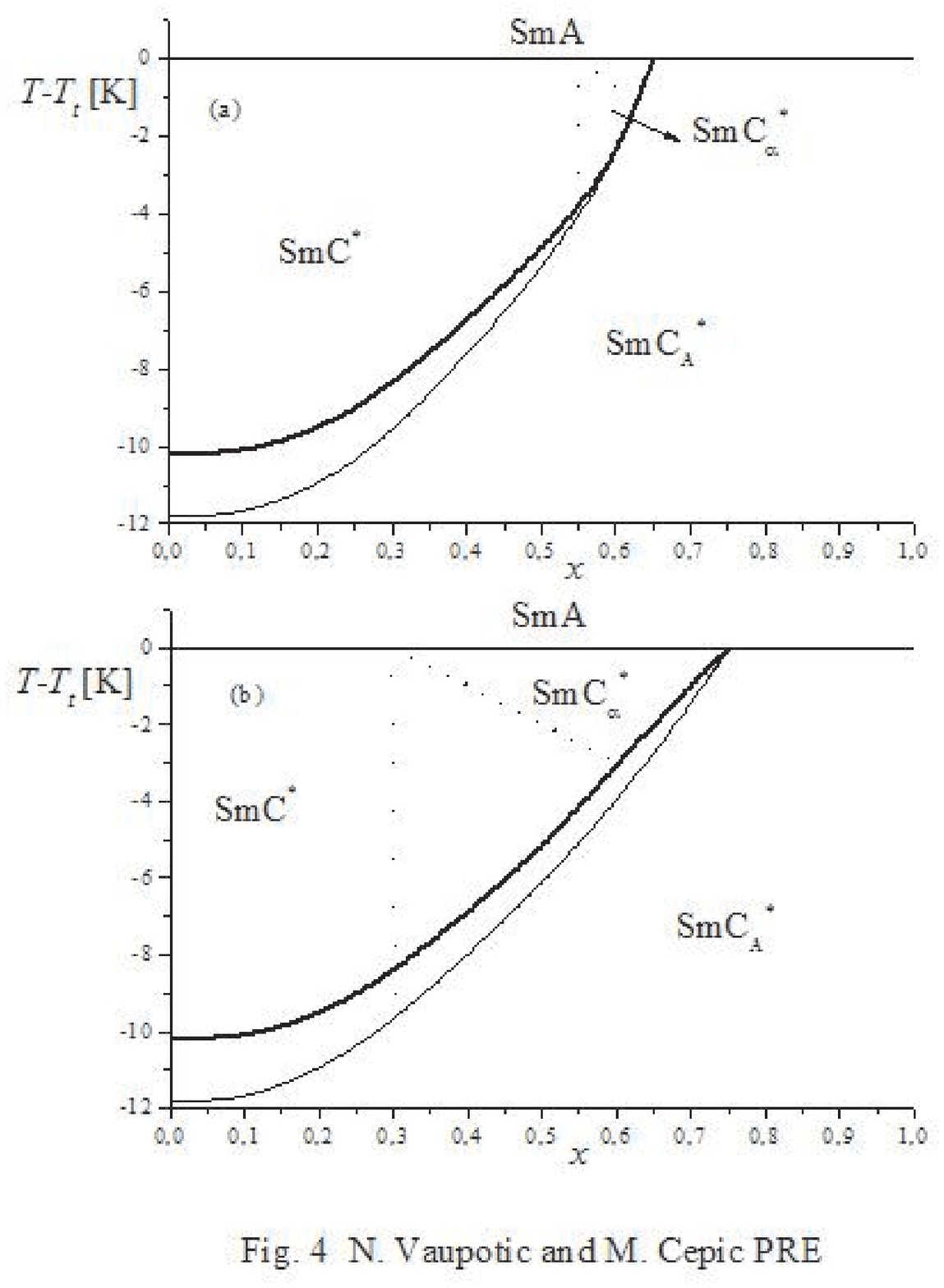}
\caption{Phase diagrams at finite flexoelectric coupling: the
effect of the sign of the effective chiral interaction and the
quadrupolar coupling. a)
$\widetilde{f}_{1}=-\widetilde{f}_{2}=0.06x$; b)
$\widetilde{f}_{1}=\widetilde{f}_{2}=-0.06x$. Both graphs: thin
solid and dotted line:  $b_{Q}=-2$; thick solid and dotted line:
$b_{Q}=-10$. Solid and dotted lines denote the first order
 and the second order transitions, respectively.
 The common set of parameters:
$\mu =0.3$, $c_{p}=4x$, $b=120$, $a_{10}=-0.4$, $a_{11}=4$, $b_{0}=2$
and $b_1 =b_0 /10$. }
\label{mi1-phd}
\end{figure}

The flexoelectric coupling has several effects. As already mentioned it
changes the effective interaction between the next nearest layers (see Eq. (%
\ref{a2v})) so that anticlinic ordering is preferred ($\widetilde{a}_{2}>0$%
). It is also important that at $\mu \ne 0$ the effective chiral
interactions up to the second (see Eq. (\ref{f2v})) and the third nearest
layers are present. This interactions have a significant effect on the
magnitude of the phase angle $\alpha $. It is shown below that when the
effective chiral interactions between the nearest and the next nearest
layers are opposing each other the region of stability of the SmC$_{\alpha
}^{*}$ phase is significantly reduced compared to the case where $\widetilde{%
f}_{1}$ and $\widetilde{f}_{2}$ have the same sign. At fixed flexo ($\mu $),
piezo ($c_{p0}$) and electrostatic ($b_{0}$, $b_{1}$) interactions the sign
of the effective chiral parameter $\widetilde{f}_{1}$ depends on the
magnitude of the direct chiral van der Waals interaction ($f_{10}$) between
the nearest layers (see Eq. (\ref{f1v})). Only minor changes in this
interaction can cause a change in the sign of $\widetilde{f}_{1}$ which then
essentially leads to a significant reduction or enlargement of the stability
region of the SmC$_{\alpha }^{\star }$ phase as shown in Fig. \ref{mi1-phd}.

In Fig. \ref{mi1-phd}a,b phase diagrams are shown for materials which differ
in the magnitude of the direct chiral van der Waals interaction ($f_{10}$).
In the phase diagram shown in Fig. \ref{mi1-phd}a the effective chiral
interactions between the nearest and the next nearest layers prefer the
opposite helicoidal modulation and are thus opposing each other. In this
case the stability of the SmC$_{\alpha }^{*}$ phase is limited to the region
of optical purity where the general condition for the stability of the SmC$%
_{\alpha }^{\star }$ phase ($4|\widetilde{a}_{1}|<\widetilde{a}_{2}$) is
satisfied. If the chiral van der Waals interaction is slightly increased
(Fig. \ref{mi1-phd}b) the effective chiral interactions favor the same sense
of the helicoidal modulation. So even in the region where the general
condition for the stability of the SmC$_{\alpha }^{\star }$ phase is not
satisfied, the helicoidal modulation is so large that the structure might be
recognized as the SmC$_{\alpha }^{\star }$ phase.

\begin{figure}
\includegraphics[width=0.8\textwidth]{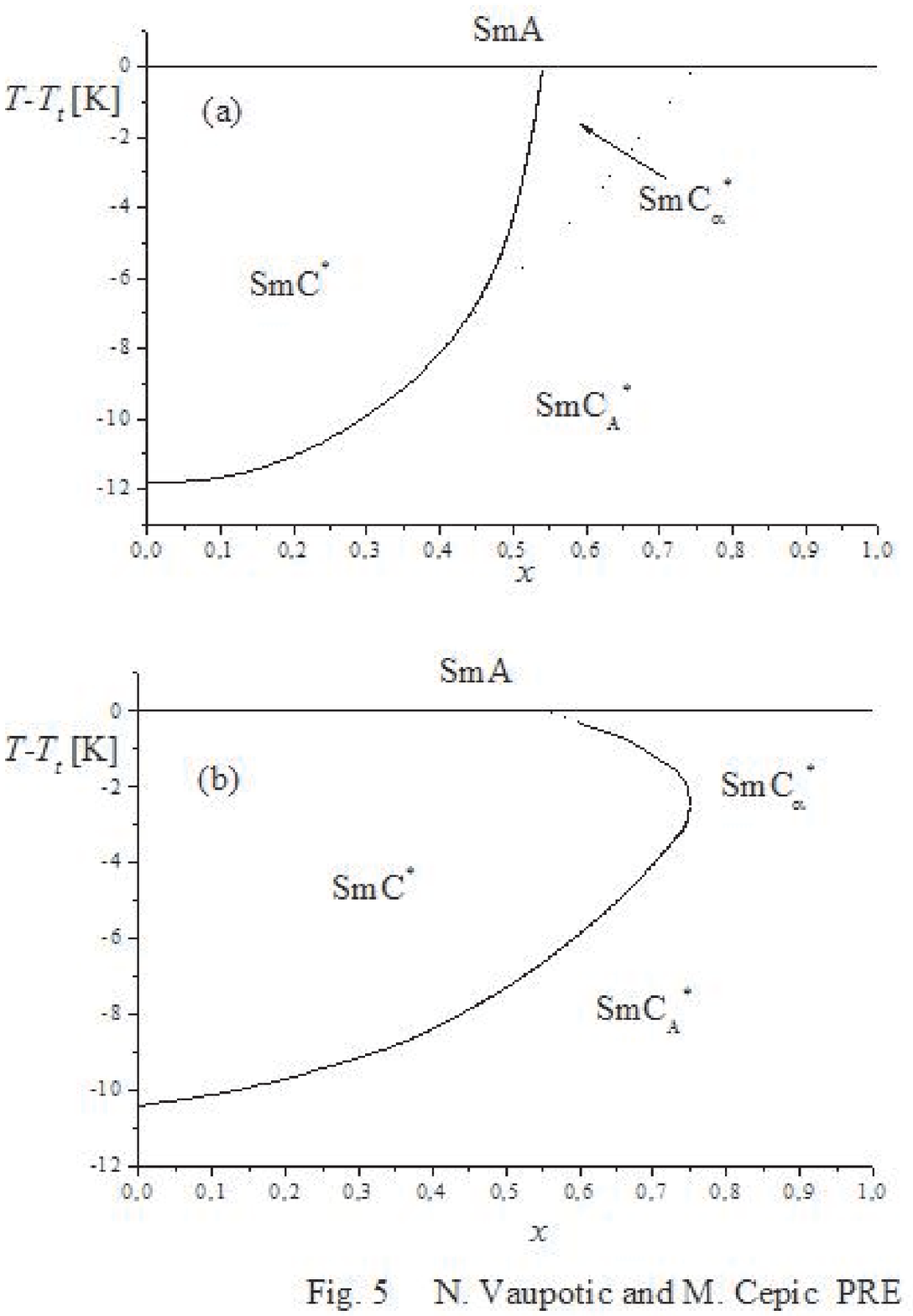}
\caption{Phase diagrams at
 $\mu =0.5$, $c_{p}=-3x$, $\widetilde{f}_{1}=-0.17x$ and $\widetilde{f}_{2}=0.075x$.
The rest of the parameters are the same as in Fig. 4.
a) $b_{Q}=-2$ and b) $b_{Q}=-10$. Solid and dotted lines denote the first order
 and the second order phase transitions, respectively. The SmC$_{\alpha }^{\star }$ phase is
designated to the region in which $\alpha < 1.5$. }
\label{mi2}
\end{figure}

\begin{figure}
\includegraphics[width=0.8\textwidth]{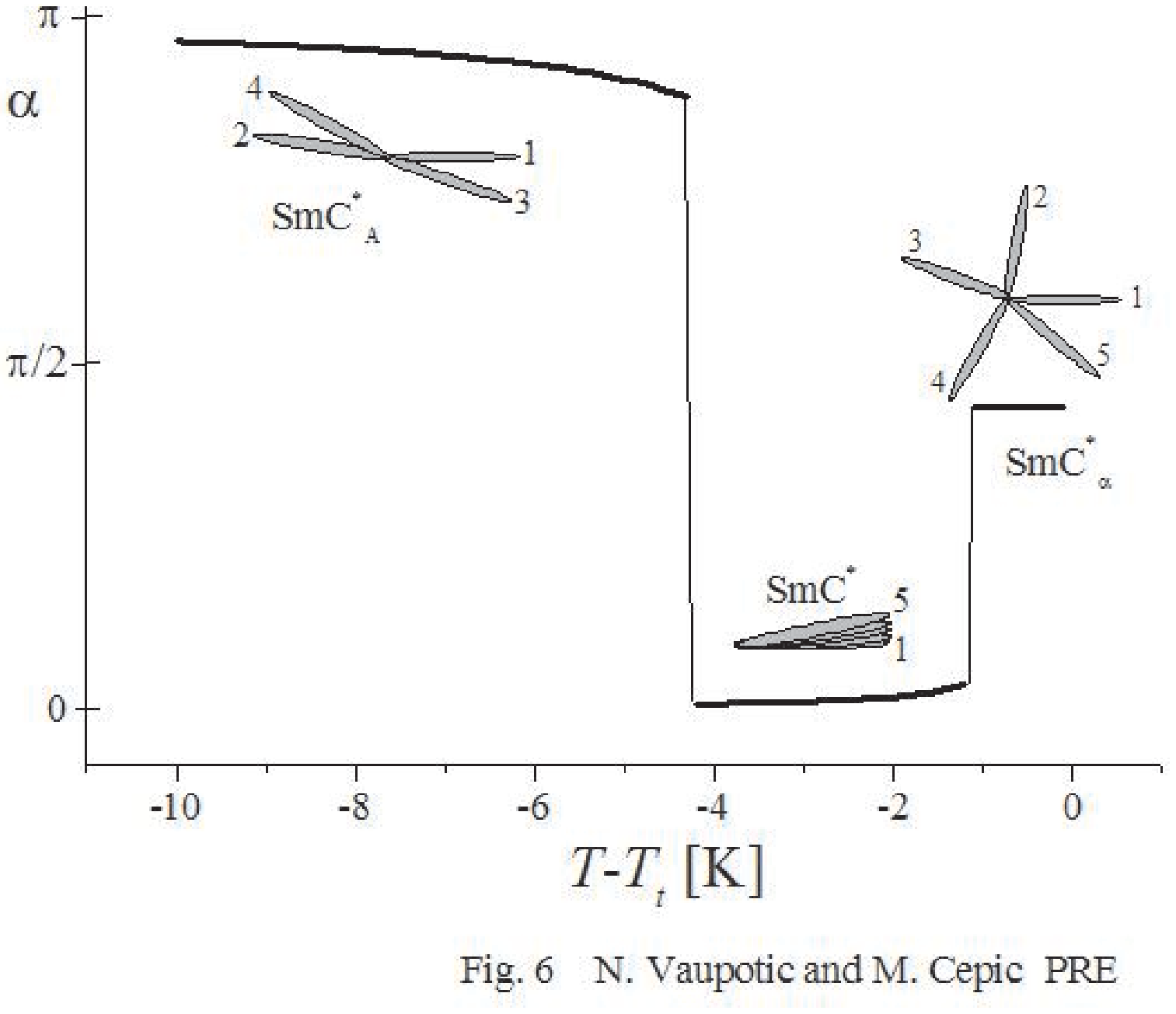}
\caption{Temperature dependence of the phase angle $\alpha $  at
$x=0.7$ for the phase diagram shown in Fig. 5b. At this optical
purity all the three clock phases are observed  when temperature
is reduced below the transition temperature $T_t $ to the tilted
phase. All the transitions are of the first order. A schematic
diagram (top view on the layers) is shown for each phase. The
numbers count the successive layers.} \label{phangle}
\end{figure}

In Fig. \ref{mi1-phd}a,b we also show the effect that the increased
quadrupolar coupling has on the phase diagram. Since quadrupolar interaction
enforces synclinic or anticlinic phases the region of stability of the SmC$%
_{\alpha }^{\star }$ phase is reduced. The SmC$^{\star }$ phase is pushed
towards higher optical purity and in a certain region of optical purity one
obtains the following phase sequence: SmA $\rightarrow $ SmC$_{\alpha
}^{\star }$ $\rightarrow $ SmC$^{\star }$ $\rightarrow $ SmC$_{A}^{\star }$
upon lowering the temperature. In the phase diagrams shown in Fig. \ref
{mi1-phd} the phase transitions SmC$^{\star }$ $\rightarrow $ SmC$%
_{A}^{\star }$ and SmC$_{\alpha }^{\star }$ $\rightarrow $ SmC$_{A}^{\star }$
are of the first order and the phase transition SmC$_{\alpha }^{\star }$ $%
\rightarrow $ SmC$^{\star }$ is continuous. However, the type of transition
depends strongly on the strength of the piezo and flexoelectric coupling and
on the effective chiral interaction between the nearest layers. In Fig. \ref
{mi2} we show the phase diagrams which are obtained if the flexoelectric
interaction and the effective chiral interaction are increased and
flexoelectric interaction is reduced. At low quadrupolar coupling one can
expect the phase diagram shown in Fig. \ref{mi2}a. The phase transition SmC$%
_{\alpha }^{\star }$ $\rightarrow $ SmC$_{A}^{\star }$ has become continuous
and the phase transition SmC$^{\star }$ $\rightarrow $ SmC$_{A}^{\star }$
remains first order. Additionally we have a first order transition SmC$%
^{\star }$ $\rightarrow $ SmC$_{\alpha }^{\star }$ at $0.45<x<0.55$. If the
quadrupolar coupling is increased the stability region of the SmC$^{\star }$
phase increases on the expense of the SmC$_{\alpha }^{\star }$ phase (see
Fig. \ref{mi2}b). The diagram is similar to the one shown in Fig. \ref
{mi1-phd}b, but the phase transition from the SmC$_{\alpha }^{\star }$ to
the SmC$^{\star }$ has become first order. Only in a small range of optical
purity the transition is still continuous. The variation of the phase angle
with temperature is shown in Fig. \ref{phangle}. At low temperatures tilts
in the neighboring layers are anticlinic and slightly distorted from the
antiparallel orientation, which is typical for the SmC$_{A}^{\star }$ phase.
Upon heating the structure discontinuously transforms into the synclinic
helicoidally modulated structure of the SmC$^{\star }$ phase. Close to the
transition temperature to the nontilted phase the phase difference again
changes discontinuously and the helicoidally modulated structure with a
short pitch, i.e. the SmC$_{\alpha }^{\star }$ phase, becomes stable.

The phase diagrams shown in Figs. \ref{mi1-phd},\ref{mi2} are again generic.
They can be stretched or compressed in the horizontal direction by a simple
rescaling of the piezoelectric parameter $c_{p0}$ and the chiral part of the
van der Waals interaction $f_{10}$. For example, if both parameters are
increased by a factor of 1.5, the 'elbow' of the SmC$^{\star }$ phase in
Fig. \ref{mi2}b would extend to $x>1$, which means, that in the optically
pure sample ($x=1$) one would observe the following phase sequence upon
lowering the temperature: SmA $\rightarrow $ SmC$_{\alpha }^{\star }$ $%
\rightarrow $ SmC$^{\star }$ $\rightarrow $ SmC$_{A}^{\star }$ and this
would agree with the schematic diagram shown in Fig. \ref{shema}c. As
already mentioned the rescaling of the diagram shown in Fig. \ref{mu0}b also
gives a phase diagram which qualitatively agrees with the one on Fig. \ref
{shema}c, however the origin of the SmC$_{\alpha }^{\star }$ phase in Figs.
\ref{mu0} and \ref{mi2} is different.

\section{Conclusions}

The discrete phenomenological model \cite{mojcaprl} reproduces the general
phase diagrams that are consistent with the experimental results \cite
{takezoe}. We have shown that the reduction of optical purity, in general,
reduces the magnitude of the inlayer polarization which forces the system
from the antiferroelectric to the ferroelectric phase with a possible SmC$%
_{\alpha }^{\star }$ phase between them.

We conclude that from the experimental phase diagrams one can predict the
most important microscopic interactions in the material. If the phase
diagram includes only two phases, SmC$^{\star }$ and SmC$_{A}^{\star }$, the
flexoelectric coupling is very small. The materials which exhibit the SmC$%
_{\alpha }^{\star }$ in a certain range of optical purity and temperature,
can be expected to have a significant direct chiral van der Waals
interaction or/and the flexoelectric coupling that is comparable with the
piezoelectric coupling. And finally, when temperature is lowered, the phase
sequence SmA $\rightarrow $ SmC$_{\alpha }^{\star }$ $\rightarrow $ SmC$%
^{\star }$ $\rightarrow $ SmC$_{A}^{\star }$ is possible only in materials
in which quadrupolar coupling is very strong. The transitions can be of the
first order only if flexoelectric coupling is significant as well. In such
materials the phases with a short modulation over three and four layers are
usually present as well \cite{mojcajcp,takezoe}. The study of these phases,
however, is already beyond the scope of this paper.

\bigskip \pagebreak

\end{document}